\documentclass{article}
\usepackage{spconf,amsmath,epsfig}
\usepackage{amssymb}
\usepackage{graphicx}
\usepackage{color}

\usepackage[mathcal]{euscript}
\usepackage{amsmath,amstext,amsfonts,amssymb}
\usepackage{shadow,color,pifont,times,rotate}
\usepackage{array}
\usepackage{epic}
\usepackage{eepic}
\usepackage{epsfig,float}
\usepackage{graphics}
\usepackage{graphicx}
\usepackage{multicol}       

\newtheorem{defi}{Definition}
\newtheorem{theo}{Theorem}
\newtheorem{prop}{Proposition}

\newcommand{\bs}{\boldsymbol}
\newcommand{\tQ}{\tilde{{\bf Q}}}
\newcommand{\tP}{\tilde{{\bf P}}}
\newcommand{\tO}{\tilde{{\bf Q}}_{*}}

\title{On the Capacity Achieving Transmit Covariance Matrices of MIMO Correlated Rician Channels: A Large
System Approach}

\name{Julien Dumont$^{12}$, Philippe Loubaton$^2$ and Samson
Lasaulce$^3$}
\address{\normalsize $^1$France Telecom R$\&$D, 38-40 Rue du General
Leclerc 92794 Issy-les-Moulineaux Cedex 9, France\\
\normalsize $^2$IGM Lab. Info., UMR-CNRS 8049, Universit{\'e} de Marne la Vall{\'e}e, 77454 Marne-la-Vall{\'e}e, France\\
\normalsize $^3$ CNRS-LSS, 5, Rue Joliot Curie, 91192 Gif-sur-Yvette, France\\
\normalsize E-mail: {\tt \{dumont,loubaton\}@univ-mlv.fr},\tt
lasaulce@lss.supelec.fr}

\begin{document}

\maketitle

\begin{abstract}
We determine the capacity-achieving input covariance matrices for
coherent block-fading correlated MIMO Rician channels. In contrast with the Rayleigh and uncorrelated Rician
cases, no closed-form expressions for the eigenvectors of the
optimum input covariance matrix are available. Both the
eigenvectors and eigenvalues have to be evaluated by using
numerical techniques. As the corresponding
optimization algorithms are not very attractive, we evaluate
the limit of the average mutual information when the number of transmit
and receive antennas converge to $+\infty$ at the same rate.
We propose an attractive optimization
algorithm of the large system approximant, and establish
some convergence results.  Numerical simulation results
show that, even for a quite moderate number of transmit and receive antennas, the new approach
provides the same results than direct maximization approaches
of the average mutual information, while being much more computationally attractive.
\end{abstract}

\section{Introduction}
Since the seminal work of Telatar (\cite{Telatar}),
it is widely recognized that the use of multiple antennas at both the
transmitter and the receiver has the potential to increase
the capacity of digital communication systems. However,
to take benefit of the potential of MIMO systems,
it is necessary to adapt the transmitter to the
channel in some optimal way. In the context of the so-called block-fading channel,
the channel matrix is generally modelled as a random complex Gaussian matrix, and
one of the most popular figure of merit is the
ergodic capacity defined as the maximum over the input covariance matrices of the average mutual information.
It is in general reasonnable to assume that the mean and the covariance of the channel are available at the transmitter side.
Therefore, the average mutual information can, in principle, be evaluated and optimized w.r.t. the input covariance
matrix at the transmitter side. \\

This optimization problem has been addressed extensively in the case of certain Rayleigh channels.
In the context of the so-called Kronecker model, it has been shown by various authors (see e.g.
\cite{Goldsmith-Jafar-etal-03} for a review) that the eigenvectors of the optimal input covariance
matrix coincide with the eigenvectors of the transmit correlation matrix. It is therefore sufficient
to evaluate the eigenvalues of the optimal matrix, a problem which can be solved by using
standard optimization algorithms. Note that \cite{Tulino-Verdu-03}  extended this result to
more general (non Kronecker) Rayleigh channels. Rician channels have been comparatively less
studied from this point of view. We mention the work \cite{Hoesli-Kim-Lapidoth-05} devoted to
the case of uncorrelated Rician channels.
\cite{Hoesli-Kim-Lapidoth-05} proved that the eigenvectors of the optimal input covariance matrix
are the right-singular vectors of the line of sight component of the channel. As in the Rayleigh case,
its eigenvalues can be evaluated by standard routines. The case of correlated Rician channels is undoubtly
more complicated because the eigenvectors of the optimum matrix have no closed form expressions.
Therefore, both its eigenvalues and its eigenvectors have to be evaluated numerically. For this, it is
necessary to use numerical methods: see in particular \cite{Vu-Paulraj-05} where
a barrier interior-point method has been implemented. The corresponding algorithms are
however not very attractive because
the exact expression of the average mutual information is quite complicated (\cite{Kang-Alouini-06}). Therefore,
its gradient and its Hessian have rather to be evaluated using computationally intensive Monte-Carlo simulation methods. \\

In this paper, we address the optimization of the input covariance of bi-correlated Rician channels.
As the exact expression of the average mutual information is quite complicated, we propose to evaluate its
limit when the number of transmit and receive antennas converge to $+\infty$ at the same rate,
and to address the optimization of its asymptotic approximation, hopefully a simpler problem.
The asymptotic expression of the mutual information has been obtained by various
authors in the case of MIMO Rayleigh channels, and has been shown to be quite reliable even for a quite moderate
number of antennas: see e.g. \cite{Chuah-Tse-02}, \cite{Tulino-Verdu-Book} in which large random matrix
results have been used, \cite{Moustakas-Simon-Sengupta-03} which uses the non rigorous, but useful,
replica method. In our knowledge, the asymptotic analysis of Rician channels has been considered in \cite{Cottatelluci-Debbah-04}
(using a result of Girko \cite{Girko-01} valid in the context of restrictive assumptions) and
\cite{Moustakas-Simon-05} (using the replica method) in the uncorrelated case
and in \cite{Dumont-Loubaton-Lasaulce-Debbah-05} in the case of receive correlated Ricean channels.
In this paper, we use the recent results of \cite{Hachem-Loubaton-Najim-05} in which a closed form
asymptotic approximation of the mutual information is provided, and state without proof new results
concerning its accuracy. Then, we address the optimization of the large system approximation w.r.t. the input covariance matrix.
As the average mutual information, the corresponding function is strictly concave. We propose a simple
iterative maximization algorithm, which, in some sense, can be seen as a generalization to the
Rician case of proposal of \cite{Wen-Com-06} devoted to the Rayleigh context: each iteration needs to solve a
system of 2 non linear equations
as well as a standard waterfilling problem. In contrast with \cite{Wen-Com-06}, we give some convergence
results: we prove that, if convergent, then the algorithm
converges toward the optimum input covariance matrix. Finally, simulation results confirm the relevance of our
approach. \\

This paper is organized as follows. Section \ref{sec:model} is devoted to the presentation of the model and of
the underlying assumptions. Section \ref{sec:asymptotic} presents our asymptotic approximation  of
the average mutual information.
Section \ref{sec:maximization} is devoted to the maximization of our mutual information approximation. Finally,
simulation results are provided in Section \ref{sec:simulation}.

\section{Presentation of the channel model}
\label{sec:model}
We consider a block fading MIMO static channel and denote by $n$ and $N$  the number of transmit and receive antennas
respectively. The $N \times n$ channel matrix, denoted ${\bs \Sigma}$, is supposed to be given by ${\bs \Sigma} = {\bf A} + {\bf Y}$.
${\bf Y}$ is a zero mean $N \times n$ complex Gaussian random matrix (sometimes called complex circular Gaussian
random matrix) given by ${\bf Y} = \frac{1}{\sqrt{n}} {\bf R}^{1/2}
{\bf X} {\bf T}^{1/2}$
where ${\bf R}$ and ${\bf T}$ are the receive and transmit correlation matrices, and where ${\bf X}$ is a zero mean independent
identically distributed complex Gaussian matrix in the sense that the real and imaginary
parts of the entries of ${\bf X}$ are independent, and have the same variance $\frac{1}{\sqrt{2}}$.
${\bf A}$ represents a deterministic $N \times n$ matrix. Very often, ${\bf A}$ is assumed to be a rank one matrix
(see e.g. \cite{Goldsmith-Jafar-etal-03}, \cite{Lozano-Tulino-Verdu-03}).
However, in important contexts, this hypothesis is not valid. Macro diversity downlink transmissions are typical examples in which ${\bf A}$
is likely to be full rank. In this context, transmit antennas are very far from each other,
while the distance between the receive antennas are of the
order of the wavelength of the transmitted signals. In such a context, the line of sight components between each transmit antenna and
the receive antenna arrays are different, so that ${\bf A}$ is likely to be full rank. If the receive antennas array is
linear and uniform, a typical example for ${\bf A}$ is
\begin{equation}
\label{eq:exempleA}
{\bf A} = \sqrt{\frac{K}{K+1}} \sqrt{\frac{N}{n}} \left[{\bf a}(\theta_1), \ldots, {\bf a}(\theta_n) \right] {\bs \Lambda}
\end{equation}
where ${\bf a}(\theta) = \frac{1}{\sqrt{N}} (1, e^{i \theta}, \ldots, e^{ i (N-1) \theta})^{T}$ and ${\bf \Lambda}$ is a diagonal
matrix, the entries of which represent the complex amplitudes of the $n$ line of sight components.
$0 < K < +\infty$ is the so-called Rice factor of the
channel. In the following, we therefore do not formulate any assumption on the rank of ${\bf A}$. Finally, matrices ${\bf A}, {\bf R}, {\bf T}$
are normalized in such a way that $\frac{1}{N} \mbox{Tr}({\bf R}) = \frac{1}{\sqrt{K+1}},\frac{1}{n} \mbox{Tr}({\bf T}) = \frac{1}{\sqrt{K+1}},
\frac{1}{N} \mbox{Tr}({\bf A}{\bf A}^{H}) = \frac{K}{K+1}$ where $0 < K < +\infty$ is the  Rice factor of the
channel.

\section{Asymptotic behaviour of the average mutual information.}
\label{sec:asymptotic}
In the following, we denote by ${\cal C}$ the cone of non negative Hermitian $n \times n$ matrices, and by
${\cal C}_1$ the subset of all matrices $\tilde{{\bf Q}}$ of ${\cal C}$ for which $\frac{1}{n} \mbox{Tr}(\tilde{{\bf Q}}) = 1$. Let $\tQ$ be an element of ${\cal C}_1$. Let $\sigma^{2}$ be a fixed noise level.
Then, we denote by $I(\tQ)$ the average mutual information at
the noise level $\sigma^{2}$ given by
\begin{equation}
\label{eq:defmutual}
I(\tilde{{\bf Q}}) =  \mathbb{E} \left[ \log \mbox{det} \left( {\bf I} + \frac{{\bs \Sigma} \tilde{{\bf Q}} {\bs \Sigma}^{H}}{\sigma^{2}} \right) \right]
\end{equation}
As it is well known, the ergodic capacity $C_E$ of the channel is defined as
\begin{equation}
\label{eq:defcapa}
C_E = \max_{\tQ \in {\cal C}_1} I(\tQ)
\end{equation}
The optimal input covariance matrix thus coincides with the argument of the above maximization problem. Note that
function $\tQ \rightarrow I(\tQ)$ is strictly concave while the set ${\cal C}_1$ on which it is defined
is convex. Therefore (\cite{Luenberger}), the maximum of $I$ on ${\cal C}_1$ is reached in a unique point.

If ${\bf R} = {\bf I}$ and ${\bf T} = {\bf I}$, it is shown in \cite{Hoesli-Kim-Lapidoth-05} that the
eigenvectors of the optimal
input covariance matrix coincides with the right-singular vectors of ${\bf A}$.  Apart this
simple case, it seems difficult to characterize in closed form the eigenvectors of the optimal matrix. Therefore, its evaluation
requires to use numerical technics (see \cite{Vu-Paulraj-05}). This approach is
complicated by the fact that the expression of function $I(\tilde{{\bf Q}})$ is quite complicated (\cite{Kang-Alouini-06}). Therefore, its gradient and Hessian have to be evaluated using Monte Carlo simulations. In the asymptotic regime $N \rightarrow +\infty$,
$n \rightarrow +\infty$ in such a way that $\frac{N}{n} \rightarrow \alpha$ where $0 < \alpha < +\infty$,
$I(\tQ)$ turns out to be equivalent to a much simpler term. The purpose of this section is to review the corresponding
asymptotic results.  In order
to simplify the notations, the symbol $n \rightarrow +\infty$ should be understood from now on as $n$ and $N$ converge
to $+\infty$ in such a way $\frac{n}{N} \rightarrow \alpha$.

$I(\tQ)$ coincides with the average mutual information of the virtual channel
$${\bs \Sigma} \tQ^{1/2} = {\bf A} \tQ^{1/2} + \frac{1}{\sqrt{n}} {\bf R}^{1/2} {\bf X} {\bf U} (\tQ^{1/2} {\bf T} \tQ^{1/2})^{1/2}$$
where matrix ${\bf U}$ is the constant $n \times n$ unitary matrix ${\bf U} = {\bf T}^{1/2} \tQ^{1/2} (\tQ^{1/2} {\bf T} \tQ^{1/2})^{-1/2}$
As ${\bf X} {\bf U}$ has the same statistical properties than ${\bf X}$, it appears that ${\bs \Sigma} \tQ^{1/2}$ can be interpreted
as a bi-correlated Gaussian Rician channel with mean ${\bf A} \tQ^{1/2}$ and receive and transmit correlation matrices
${\bf R}$ and $\tQ^{1/2} {\bf T} \tQ^{1/2}$ respectively. In the following, we denote by ${\bf T}(\tQ)$ the matrix
${\bf T}(\tQ) = \tQ^{1/2} {\bf T} \tQ^{1/2}$.
In order to derive an asymptotic approximation
of $I(\tQ)$, it is therefore possible to use the results of \cite{Hachem-Loubaton-Najim-05}. We note that
the results of \cite{Hachem-Loubaton-Najim-05} are obtained if matrices ${\bf R}$ and $\tQ^{1/2} {\bf T} \tQ^{1/2}$
are diagonal.
The unitary invariance of the mutual information of Gaussian random matrices allows however to use these results.
We first state the following result, which derives
partly from \cite{Hachem-Loubaton-Najim-05}.
\begin{theo}
\label{theo:canonique}
Assume that $\sup_{n} \|{\bf A}\| < +\infty$,  $\sup_{n} \|{\bf R}\| < +\infty$,  $\sup_{n} \|{\bf T}\| < +\infty$, and
$\sup_{n} \|\tQ \| < +\infty$
where $\| . \|$ stands for the spectral norm. Consider the system of equations
\begin{equation}
\label{eq:canonique}
\left\{
\begin{array}{ccc}
\kappa & = & f(\kappa, \tilde{\kappa}, \tQ)  \\
\tilde{\kappa} & = & \tilde{f}(\kappa, \tilde{\kappa}, \tQ)
\end{array}\right. .
\end{equation}
where $f(\kappa, \tilde{\kappa}, \tQ)$ is given by
\begin{equation}
\label{eq:deff}
\frac{1}{n} \textrm{Tr} \left[ {\bf R} \left( \sigma^{2}({\bf I}+{\bf R} \tilde{\kappa}) + {\bf A} \tQ^{1/2}({\bf I}+{\bf T}(\tQ) \kappa)^{-1}\tQ^{1/2} {\bf A}^{H} \right)^{-1} \right]
\end{equation}
and $\tilde{f}(\kappa, \tilde{\kappa}, \tQ)$ by
\begin{equation}
\label{eq:deftildef}
\frac{1}{n} \textrm{Tr} \left[ {\bf T}(\tQ) \left( \sigma^{2}({\bf I}+ {\bf T}(\tQ) \kappa) + \tQ^{1/2} {\bf A}^{H}
({\bf I}+ {\bf R} \tilde{\kappa})^{-1} {\bf A} \tQ^{1/2} \right)^{-1} \right]
\end{equation}
Then, equations (\ref{eq:canonique}) have unique strictly positive solutions $(\delta(\tQ), \tilde{\delta}(\tQ))$.
Moreover, when $n \rightarrow +\infty$,
\begin{equation}
\label{eq:equivalent1}
I(\tQ) = \overline{I}(\tQ) + O(1/n)
\end{equation}
where the asymptotic approximant $\overline{I}(\tQ)$ is defined by
\begin{equation}
\label{eq:expreCbarre}
\begin{array}{c}
\overline{I}(\tQ) = \log \mbox{det} \left[{\bf I} + {\bf H}(\tQ) \tQ {\bf H}(\tQ) \right]  \\
+  \log \mbox{det} \left[{\bf I} + \tilde{\delta}(\tQ) {\bf R} \right] - \sigma^{2} n \delta(\tQ) \tilde{\delta}(\tQ)
\end{array}
\end{equation}
where ${\bf H}(\tQ)$ represents the $n \times n$ positive definite \footnote{${\bf H}(\tQ)$ is positive definite
because $\delta(\tQ) > 0$} matrix defined by
\begin{equation}
\label{eq:defH}
{\bf H}(\tQ) = \left[ \delta(\tQ) {\bf T} + \frac{1}{\sigma^{2}} {\bf A}^{H}({\bf I} + \tilde{\delta}(\tQ) {\bf R})^{-1} {\bf A}  \right]^{1/2}
\end{equation}
\end{theo}
The proof of this result is far from being obvious, and is of course omitted. It is partly based on
the results of \cite{Hachem-Loubaton-Najim-05}, from which one can deduce that $I(\tQ) = \overline{I}(\tQ) + o(n)$.
The fact that $I(\tQ) - \overline{I}(\tQ) = O(1/n)$ is not obvious at all, and follows specifically from the fact that matrix
${\bs \Sigma}$ has a
Gaussian complex distribution. In particular, in the Gaussian real case, $I(\tQ) - \overline{I}(\tQ) = O(1)$.
This in accordance with \cite{Bai-Silverstein-04} in which a similar result is proved
in the simpler context ${\bf A} = 0$ and $\tilde{{\bf Q}} = {\bf I}$, and with the predictions
of the replica method in \cite{Moustakas-Simon-Sengupta-03} in the case ${\bf A} = 0$ and \cite{Moustakas-Simon-05}
in the case ${\bf R} = {\bf I}$, ${\bf T} = {\bf I}$ and $\tilde{{\bf Q}} = {\bf I}$.
This very fast convergence rate tends to explain why the asymptotic evaluations of the
mean mutual information are reliable even for a quite moderate number of antennas, as remarked e.g.
in \cite{Biglieri-Taricco-Tulino-02}. See Section \ref{sec:simulation} for simulation evidence. \\

We end this section by a very useful remark. Consider the function $V(\kappa, \tilde{\kappa}, \tQ)$ defined by
replacing in (\ref{eq:expreCbarre}) solutions $(\delta(\tQ), \tilde{\delta}(\tQ))$ of (\ref{eq:canonique}) by fixed parameters
$(\kappa, \tilde{\kappa})$:
\begin{equation}
\label{eq:expreV}
\begin{array}{c}
V(\kappa, \tilde{\kappa}, \tQ) = \log \mbox{det} \left[{\bf I} + {\bf G}(\kappa, \tilde{\kappa}) \tQ {\bf G}(\kappa, \tilde{\kappa}) \right]  \\
+  \log \mbox{det} \left[{\bf I} + \tilde{\kappa} {\bf R} \right] - \sigma^{2} n \kappa \tilde{\kappa}
\end{array}
\end{equation}
where ${\bf G}(\kappa, \tilde{\kappa})$ represents the $n \times n$ positive definite  matrix defined by
\begin{equation}
\label{eq:defH}
{\bf G}(\kappa, \tilde{\kappa}) = \left[ \kappa {\bf T} + \frac{1}{\sigma^{2}} {\bf A}^{H}({\bf I} +  \tilde{\kappa} {\bf R})^{-1} {\bf A} \right]^{1/2}
\end{equation}
We of course note that ${\bf H}(\tQ) = {\bf G}(\delta(\tQ), \tilde{\delta}(\tQ), \tQ)$ and $I(\tQ) = V(\delta(\tQ), \tilde{\delta}(\tQ), \tQ)$. It is straightforward to check that
\begin{eqnarray}
\label{eq:derivees}
\frac{\partial V}{\partial \tilde{\kappa}} & = & -n \sigma^{2} \left( \kappa - f(\kappa, \tilde{\kappa}, \tQ) \right)  \nonumber \\
\frac{\partial V}{\partial \kappa} & = & -n \sigma^{2} \left( \tilde{\kappa} - \tilde{f}(\kappa, \tilde{\kappa}, \tQ) \right)
\end{eqnarray}
As $(\delta(\tQ), \tilde{\delta}(\tQ))$ satisfy Eq. (\ref{eq:canonique}), we get immediately that
\begin{eqnarray}
\label{eq:annulation-derivees}
\left( \frac{\partial V}{\partial \kappa} \right)_{(\delta(\tQ), \tilde{\delta}(\tQ), \tQ)} & = & 0 \nonumber \\
\left( \frac{\partial V}{\partial \tilde{\kappa}} \right)_{(\delta(\tQ), \tilde{\delta}(\tQ), \tQ)} & = & 0
\end{eqnarray}
This simple observation is the key point of our input covariance  optimization algorithm.

\section{The input covariance optimization algorithm.}
\label{sec:maximization}
The results of Section \ref{sec:asymptotic} show that $I(\tQ)$ can be approximated with a good accuracy
by $\overline{I}(\tQ)$. Therefore, the optimum input covariance matrix can itself be approximated
by the argument of the maximum of $\overline{I}(\tQ)$ over the set ${\cal C}_1$. In this section, we propose an
attractive maximization algorithm of $\overline{I}(\tQ)$. Before presenting the algorithm, we have to introduce
some concepts and results.
\begin{defi}
\label{def:Gateau}
Let $W(\tQ)$ be a function defined on ${\cal C}_1$. If $\tQ, \tP$ are 2 elements of ${\cal C}_1$,
then $W$ is said to be differentiable in the Gateaux sense at point $\tQ$ in the
direction $\tP - \tQ$ if the limit
\begin{equation}
\label{eq:defGateau}
\lim_{\lambda \rightarrow 0^{+}} \frac{W \left( \tQ + \lambda(\tP - \tQ) \right) - W(\tQ)}{\lambda}
\end{equation}
exists. In this case, this limit is denoted $<W'(\tQ), \tP - \tQ>$.
\end{defi}
Note that for each $\lambda \in [0,1]$, matrix $ \tQ + \lambda(\tP - \tQ) = (1-\lambda) \tQ + \lambda \tP$
of course belongs to ${\cal C}_1$. Therefore, \\ $W \left( \tQ + \lambda(\tP - \tQ) \right)$ makes sense for
$\lambda > 0$ small enough.
\begin{prop}
\label{prop:caracterisation}
Let $W$ be a strictly concave function defined on ${\cal C}_1$. Then, the maximum of
$W$ on ${\cal C}_1$ is reached at a unique point $\tO$ of ${\cal C}_1$. Assume that for every elements $\tQ, \tP$
of ${\cal C}_1$, $W$ is differentiable in the Gateaux sense at point $\tQ$ in the direction
$\tP - \tQ$. Then, $\tO$ is the unique element of ${\cal C}_1$ verifying
\begin{equation}
\label{eq:condition-opt}
<W'(\tO), \tQ - \tO> \, \leq 0
\end{equation}
for each element $\tQ$ of ${\cal C}_1$.
\end{prop}
This result is a simple adaptation of known results (see e.g. \cite{Luenberger}). The proof is therefore omitted.
We now give some useful properties of function $\overline{I}$.


\begin{prop}
\label{prop:proprietesIbarre}
Function $\overline{I}(\tQ)$ is strictly concave on ${\cal C}_1$. Moreover, for every elements
$\tQ, \tP$ of ${\cal C}_1$, $\overline{I}$ is differentiable in the Gateaux sense
at point $\tQ$ in the direction $\tP - \tQ$.
\end{prop}
The fact that $\overline{I}$ is Gateaux differentiable is rather obvious. The strict concavity of
$\overline{I}$ needs some work, but is not surprising because it is an approximant of a strictly concave
function. \\

Proposition \ref{prop:caracterisation} thus implies that the maximum of $\overline{I}$ on ${\cal C}_1$ is
reached at a unique point denoted $\tilde{{\bf Q}}_{*}$. Before presenting our maximization algorithm
of $\overline{I}$, we first give some insights on the structure of matrix $\tilde{{\bf Q}}_{*}$. For this, we denote $\delta(\tilde{{\bf Q}}_{*})$
and  $\tilde{\delta}(\tilde{{\bf Q}}_{*})$ by $\delta_{*}$ and
$\tilde{\delta}_{*}$ respectively. Then, we have the following result.

\begin{prop}
\label{prop:waterfilling}
Matrix $\tilde{{\bf Q}}_{*}$ is the solution of the standard Water-Filling problem:
Maximize over $\tilde{Q} \in {\cal C}_1$ the function
\[
U(\tQ) =  \log \mbox{det} \left[{\bf I} + {\bf G}(\delta_{*}, \tilde{\delta}_{*}) \tQ {\bf G}(\delta_{*}, \tilde{\delta}_{*}) \right]
\]
where $ {\bf G}(\delta_{*}, \tilde{\delta}_{*}) =  \left[ \delta_{*} {\bf T} + \frac{1}{\sigma^{2}} {\bf A}^{H}({\bf I} +  \tilde{\delta}_{*} {\bf R})^{-1} {\bf A} \right]^{1/2}$.
\end{prop}
{\bf Proof.} The proof of this result is based on the following identity, to be proved below:
\begin{equation}
\label{eq:egalite}
<\overline{I}'(\tO), \tQ -  \tO> = <V' \left( \delta_{*}, \tilde{\delta}_{*},\tO \right),  \tQ -  \tO>
\end{equation}
for each $\tQ \in {\cal C}_1$, where  $ <V'(\delta_{*}, \tilde{\delta}_{*}, \tO), \tQ -  \tO>$
represents the Gateaux differential of function  $\tQ \rightarrow V(\delta_{*}, \tilde{\delta}_{*}, \tQ)$. In effect, if (\ref{eq:egalite}) holds,
then, Proposition \ref{prop:caracterisation} implies that
$$ <V' \left( \delta_{*}, \tilde{\delta}_{*},\tO \right),  \tQ -  \tO> \leq 0$$
for each $\tQ \in {\cal C}_1$. By Proposition \ref{prop:caracterisation}, $\tO$ maximizes the function  $\tQ \rightarrow V(\delta_{*}, \tilde{\delta}_{*}, \tQ)$,
i.e. $U(\tQ)$ because the latter functions differ up to a constant term. It remains
to prove (\ref{eq:egalite}). For this, we remark that, by (\ref{eq:annulation-derivees}),
\begin{eqnarray}
\label{eq:crucial}
\left( \frac{\partial V}{\partial \kappa} \right)_{(\delta_{*}, \tilde{\delta}_{*}, \tO)} & = & 0 \nonumber \\
\left( \frac{\partial V}{\partial \tilde{\kappa}} \right)_{(\delta_{*}, \tilde{\delta}_{*}, \tO)} & = & 0
\end{eqnarray}
On the other hand, for each $\tQ, \tP$,
\begin{equation}
\begin{array}{c}
<\overline{I}'(\tP), \tQ -  \tP> = <V'(\delta(\tP), \tilde{\delta}(\tP), \tP), \tQ -\tP)> + \\
\left( \frac{\partial V}{\partial \kappa} \right)_{(\delta(\tP), \tilde{\delta}(\tP), \tP)} < \delta'(\tP), \tQ -  \tP> + \\
\left( \frac{\partial V}{\partial \tilde{\kappa}} \right)_{(\delta(\tP), \tilde{\delta}(\tP), \tP)} < \tilde{\delta}'(\tP), \tQ -  \tP>
\end{array}
\end{equation}
where  $< \delta'(\tP), \tQ -  \tP>$ and $< \tilde{\delta}'(\tP), \tQ -  \tP>$ represent the Gateaux differentials of
functions $\delta$ and $\tilde{\delta}$. Eq. (\ref{eq:crucial}) thus implies (\ref{eq:egalite}). \\

$\delta_{*}$ and $\tilde{\delta}_*$ depend on matrix $\tilde{{\bf Q}}_{*}$. Therefore, Proposition \ref{prop:waterfilling} does not provide
by itself any optimization algorithm. However, it gives insights on the structure of $\tilde{{\bf Q}}_{*}$. Consider first the case ${\bf R} = {\bf I}$
and ${\bf T} = {\bf I}$. Then,  $ {\bf G}(\delta_{*}, \tilde{\delta}_{*})$ is a linear combination of ${\bf I}$ and matrix ${\bf A}^{H} {\bf A}$.
The eigenvectors of $\tilde{{\bf Q}}_{*}$ thus coincide with the right singular vectors of matrix ${\bf A}$, a result consistent
with the work \cite{Hoesli-Kim-Lapidoth-05} devoted to the maximization of the average mutual information $I({\bf \tQ})$. If ${\bf R} = {\bf I}$
and ${\bf T} \neq {\bf I}$,  $ {\bf G}(\delta_{*}, \tilde{\delta}_{*})$ can be interpreted as a linear combination of matrices ${\bf T}$ and
${\bf A}^{H} {\bf A}$. Therefore, if the transmit antennas are correlated, the eigenvectors of the optimum matrix $\tilde{{\bf Q}}_{*}$
coincide with the eigenvectors of some weighted sum of ${\bf T}$ and ${\bf A}^{H} {\bf A}$. This result provides a simple explanation
of the impact of correlated transmit antennas on the structure of the capacity-achieving input covariance matrix. The effect of correlated
receive antennas on  $\tilde{{\bf Q}}_{*}$ is however less intuitive because matrix  ${\bf A}^{H} {\bf A}$ has to be replaced
by  ${\bf A}^{H}({\bf I} +  \tilde{\delta}_{*} {\bf R})^{-1} {\bf A}$. \\

We are now in position to introduce our maximization algorithm of $\overline{I}$. It is mainly motivated
by the simple observation that for each fixed $(\kappa, \tilde{\kappa})$, the maximization w.r.t.
$\tQ$ of function $V(\kappa, \tilde{\kappa}, \tQ)$ defined by (\ref{eq:expreV}) can be achieved by a standard
Waterfilling procedure, which,
of course, does not need the use of numerical technics. On the other hand,
for $\tQ$ fixed, the equations (\ref{eq:canonique}) have unique solutions that, in practice,
can be obtained using a standard fixed-point algorithm. Our algorithm thus consists in adapting
parameters $\tQ$ and $\delta, \tilde{\delta}$ separately by the following iterative scheme:
\begin{itemize}
\item Initialization: $\tQ_{0} = {\bf I}$, $(\delta_1, \tilde{\delta}_1)$ are defined as the unique
solutions of system (\ref{eq:canonique}) in which $\tQ = \tQ_{0} = {\bf I}$.
Then, define $\tQ_1$ are the maximum of function $\tQ \rightarrow V(\delta_1, \tilde{\delta}_1, \tQ)$
on ${\cal C}_1$.
\item Iteration $k$: assume $\tQ_{k-1}$, $(\delta_{k-1}, \tilde{\delta}_{k-1})$ available. Then,
 $(\delta_k, \tilde{\delta}_k)$ is defined as the unique solution of (\ref{eq:canonique}) in which $\tQ = \tQ_{k-1}$.
Then, define $\tQ_k$ are the maximum of function $\tQ \rightarrow V(\delta_{k}, \tilde{\delta}_{k}, \tQ)$
on ${\cal C}_1$.
\end{itemize}
We now study the convergence properties of this algorithm, and state a result, which implies that if the
algorithm converges, then it converges to the global maximum of $\tilde{I}$.
\begin{prop}
\label{eq:convergence}
Assume that the 2 sequences $(\delta_k)_{k \geq 0}$ and  $(\tilde{\delta}_k)_{k \geq 0}$ verify
\begin{equation}
\label{eq:condition}
\lim_{k \rightarrow +\infty} \delta_k - \delta_{k-1} \rightarrow 0, \lim_{k \rightarrow +\infty} \tilde{\delta}_k - \tilde{\delta}_{k-1} \rightarrow 0
\end{equation}
Then, the sequence $(\tQ_{k})_{k \geq 0}$ converges toward the maximum $\tO$ of $\overline{I}$ on ${\cal C}_1$.
\end{prop}
Due to the lack of space, the proof is omitted.

Proposition \ref{eq:convergence} implies that if the sequence $(\tQ_k)_{k \geq 0}$ is convergent, then,
its limit coincides with the optimum matrix $\tO$. In fact, if $(\tQ_k)_{k \geq 0}$ converges,
then the 2 sequences $(\delta_k)_{k \geq 0}, (\tilde{\delta}_k)_{k \geq 0}$ also converge. This of course implies
condition (\ref{eq:condition}), and the convergence of  $(\tQ_k)_{k \geq 0}$ toward $\tO$.

Unfortunately, we have not been able to prove the convergence of $(\tQ_k)_{k \geq 0}$ by itself. However, all the numerical
experiments we have conducted tend to indicate that the algorithm is convergent. In any case, condition
(\ref{eq:condition}) is very easy to verify during the algorithm execution. In case of non convergence, other numerical
technics could be used in order to optimize $\overline{I}(\tQ)$, a simpler task
than the optimization of $I(\tQ)$.

\section{Comparison with the Vu-Paulraj's algorithm.}
\label{sec:simulation} 
\begin{figure}[ht]
\centerline{\includegraphics[scale=0.6]{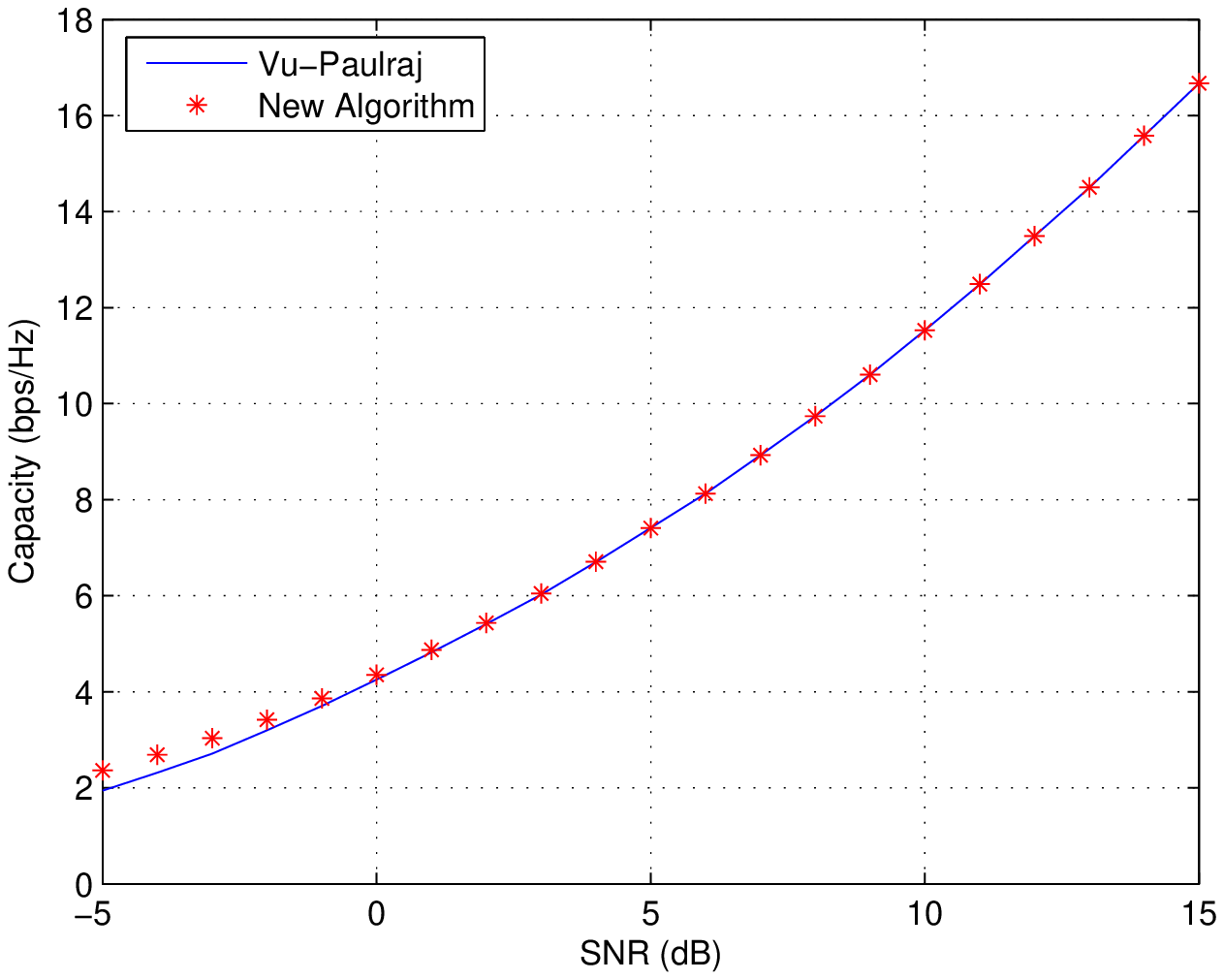}}\caption{Comparison with
the Vu-Paulraj's algorithm I} \label{fig:canal-paulraj}
\end{figure}

In this section, we compare our algorithm with the
method presented in \cite{Vu-Paulraj-05} based on the maximization of $I(\tQ)$.
We recall that Vu-Paulraj's algorithm is based on a Newton method and a barrier
interior point method. Moreover, the average mutual informations and their
first and second derivatives are evaluated by Monte-Carlo simulations. In fig.
\ref{fig:canal-paulraj}, we have evaluated $C_E = \max_{\tQ \in {\cal C}_1}
I(\tQ)$ versus the SNR for $n=N=4$. Matrix ${\bf H}$ coincides with the example
considered in \cite{Vu-Paulraj-05}. The solid line corresponds to the results
provided by the Vu-Paulraj's algorithm; the number of trials used to evaluate
the mutual informations and its first and second derivatives is equal to
$30.000$, and the maximum number of iterations is fixed to 10. The dashed line
corresponds to the results provided by our algorithm: each point represent
$I(\tO)$ at the corresponding SNR, where $\tO$ is the "optimal" matrix provided
by our approach; the average mutual information at point $\tO$ is evaluted by
Monte-Carlo simulation (30.000 trials are used). The number of iterations is
also limited to 10. Figure \ref{fig:canal-paulraj} shows that our asymptotic
approach provides the same results than the Vu-Paulraj's algorithm. However,
our algorithm is computationally much more efficient as the above table  shows.
The table gives the average executation time (in sec.) of one iteration for
both
algorithms for $n=N=2, n=N=4, n=N=8$. \\

\begin{figure}[ht]
\centerline{\includegraphics[scale=0.6]{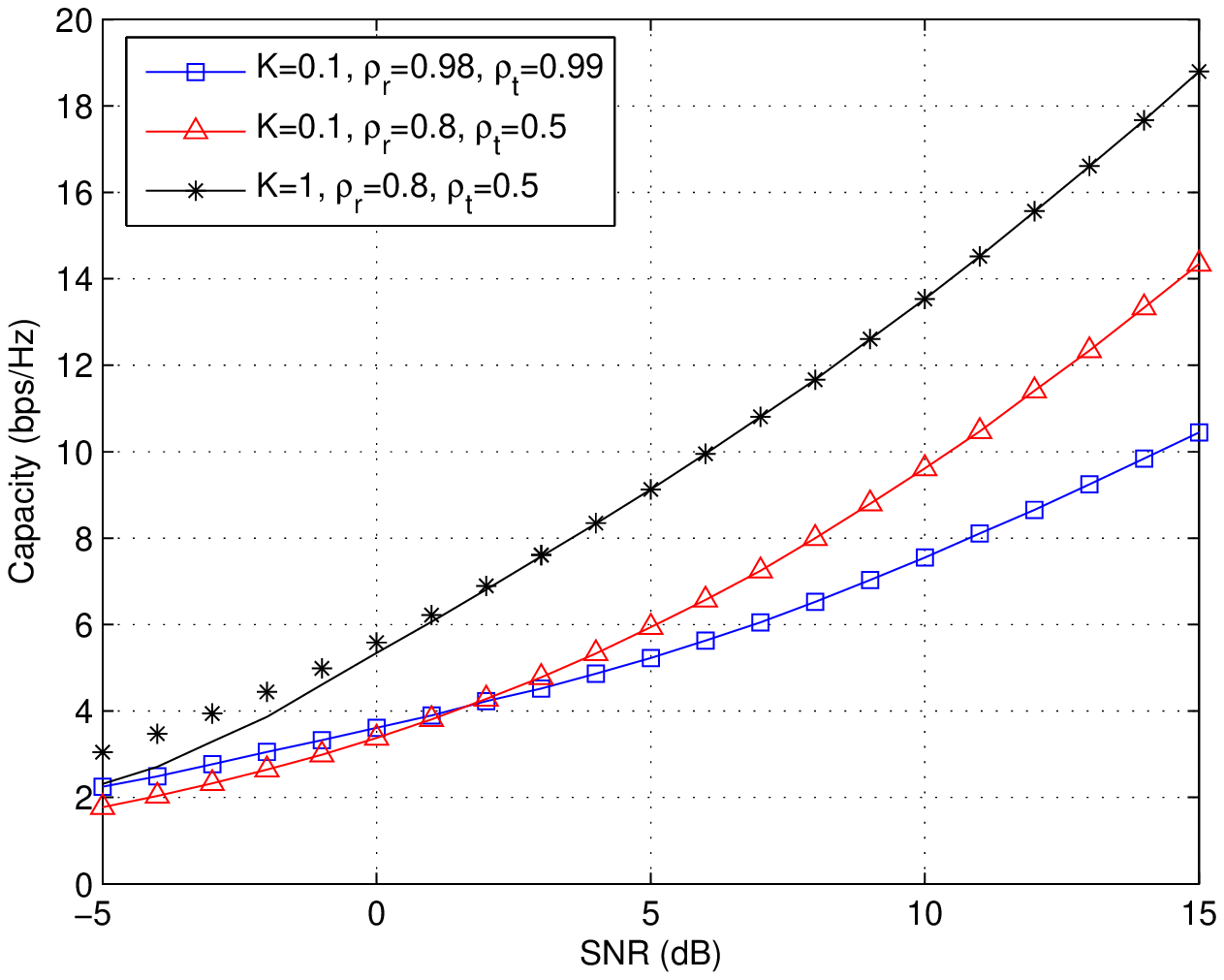}} \caption{Comparison with
the Vu-Paulraj's algorithm II} \label{fig:canal-aleatoire}
\end{figure}

In fig. \ref{fig:canal-aleatoire}, we again compare Vu-Paulraj's algorithm and our proposal.
Matrix ${\bf A}$ is generated according to (\ref{eq:exempleA}),
the angles being chosen at random. The transmit and receive antennas correlations are exponential with parameter
$0 < \rho_{t} < 1$ and $0 < \rho_{r} < 1$ respectively. In the experiments, $n=N=4$, while various values of
$\rho_t$, $\rho_r$ and of the Rice factor $K$ have been considered. As in the previous experiment,
the maximum number of iterations for both algorithms is 10, while the number of trials generated to
evaluate the average mutual informations and their derivatives is equal to 30.000. Our approach again provides the same
results than Vu-Paulraj's algorithm, except for low SNRs for $K=1, \rho_t=0.5, \rho_r=0.8$ where our method
gives better results: at these points, the Vu-Paulraj's algorithm seems not to have converge
at the 10th iteration.

\begin{figure}
\label{table}
\begin{center}
\begin{tabular}{c|c|c|c|}
\cline{2-4}
& $n=N=2$ & $n=N=4$ & $n=N=8$ \\
\hline
\multicolumn{1}{|c|}{Vu-Paulraj} & $0.75$ & $8.2$& $138$\\
\hline
\multicolumn{1}{|c|}{New algorithm} & $10^{-2}$ & $3.10^{-2}$ & $7.10^{-2}$ \\
\hline
\end{tabular}
\end{center}
\caption{Average time per iteration in seconds}
\end{figure}

\section{Conclusion}

In this paper we proposed a new approach to characterize the
capacity achieving covariance matrix of bi-correlated Rician
MIMO channels. We proposed to approximate the average mutual
information by its large system limit and derived an attractive
iterative optimization algorithm which does not need the use of
intricate numerical techniques. We have shown that the
algorithm (when it is convergent) converges to the maximum of the
approximate mutual information. Numerical simulation results
show that the new approach provides the same results than direct
maximization approaches
of the mutual information, while being much more
computationally attractive.

\end{document}